\newcommand{\beq}{\begin{equation}}
\newcommand{\eeq}{\end{equation}}

\newcommand{\bear}{\begin{eqnarray}}
\newcommand{\eear}{\end{eqnarray}}

\newcommand{\tn}{\textnormal}
\documentclass{JINST}
\usepackage{graphicx}
\usepackage{amssymb}
\usepackage{array}
\title{Detectors and Concepts for sub-100 ps timing with gaseous detectors}

\author{D. Gonz\'alez-D\'iaz,$^{a,b}$\thanks{Corresponding author.} ~ F. R.~Palomo,$^{c}$ J.~González,$^{a,d}$ Huangshan Chen,$^{e}$\thanks{Presently at Heidelberg University.}

\\
\llap{$^{a}$}
CERN, Geneva, Switzerland\\
\llap{$^{b}$}
Department of Physics, Uluda\u{g} University, Bursa, Turkey\\
\llap{$^{c}$}
Escuela Superior de Ingenieros, Universidad de Sevilla, Spain\\
\llap{$^{d}$}
CIEMAT, Madrid, Spain\\
\llap{$^{e}$}
Department of Engineering Physics, Tsinghua University, Beijing, China\\

E-mail: \email{Diego.Gonzalez.Diaz@cern.ch}
}

\abstract{We give a short compendium of the main ongoing detectors and concepts capable of performing accurate sub-100 ps timing at high particle fluxes and on large areas, through technologies based on gaseous media. We briefly discuss the state-of-the-art, technological limitations and prospects, and a new bizarre idea.}


\begin{document}

\section{Introduction}\label{Introduction}

Contemporary gaseous detectors relying on the resistive plate chamber (RPC) technique \cite{Pestov, Santonico} provide already, quite effortlessly, time accuracies around 100~ps-$\sigma$ for both ionizing and non-ionizing radiation thanks to the multi-gap concept \cite{Williams, Fonte}. Their range of demonstrated applicability encompasses the minimum ionizing particles (mips) characteristic of high energy particle physics (e.g.: ALICE at LHC \cite{ALICE}, STAR at RHIC \cite{STAR}, or HADES at SIS \cite{HADES}), $\gamma$-rays \cite{PET}, relativistic neutrons \cite{Neutrons} and heavy ion fragments up to $Z$=54 \cite{Caba,Para}. They are employed in 100m$^2$-size systems as long as the flux of incoming particles does not exceed 1 kHz/cm$^2$ mip-equivalent ($\sim 2$ nA/cm$^2$), during live-times corresponding to several years and to 10's of mC/cm$^2$, at least. The widespread commercial use of float glass, that is commonly used to make the resistive plates, together with its relatively good radio-frequency behaviour and good understanding of the signal propagation characteristics \cite{DiegoIntegrity, FonteProp}, pushes the maximum realizable module sizes beyond the demand of any present or future experiment, reaching 2 m of length (e.g., \cite{R3B}). With the introduction of Chinese glass in \cite{WangChinese}, detectors can be operated without degradation of timing performance up to $\sim 20$ kHz/cm$^2$, albeit at the price of a reduced module size down to $30$cm $\times$ $30$cm.

While it seems clear that RPCs can increasingly approach and establish the 100 kHz/cm$^2$ (mip-equivalent) landmark in small detectors \cite{Laso}, it is at present doubtful that they can reach any time soon the rate capability that Micro-Pattern Gaseous Detectors (MGPDs) intrinsically provide, well above 10~MHz/cm$^2$ \cite{Fil}, with detector sizes already at the m$^2$-scale \cite{CMS}. On the negative side, proportional detectors based on wires or MPGDs present, historically, a strong limitation for timing related to the physical jitter stemming from the collection time of the primary ionization. To avoid this limitation, while preserving their characteristic high rate behaviour, an intriguing scheme has been recently resurrected and is later discussed.

We will briefly present here some of the most fashionable ideas for the sub-100 ps detection of charged particles with gaseous detectors at high particle flux (section \ref{HighFlux}), and large areas (section \ref{LargeArea}) together with the present technological limits (section \ref{Horizon}). In section \ref{LargeArea} we elaborate on the concept of `compensation', an idea that holds the promise for virtually unlimited detector sizes. For a recent overview on next-generation timing detectors the reader is referred to \cite{Sebastian}, while an overview on the RPC field can be found in \cite{DiegoArchana}.

\section{Timing at high particle flux}\label{HighFlux}

\subsection{Timing-RPCs}\label{RPCs}

The time resolution of Resistive Plate Chambers can be approximated through \cite{PauloTime}, as:
\beq
\sigma_t(E) = \frac{1}{\sqrt{N}} \sqrt{\frac{\lambda_0}{g}} \frac{M}{(\alpha(E)-\eta(E))v_d(E)} \label{TimeRPC}
\eeq
where $N$ represents the number of gas gaps and $g$ their size, $\lambda_0$ is the mean free path of the impinging particle before an ionizing encounter, $\alpha(E)$ and $\eta(E)$ are
the multiplication and attachment coefficients, $v_d(E)$ is the drift velocity and $M$ is a factor of order 1
that accounts for the avalanche statistics, and for the fact that not all primary ionizations contribute equally. It turns out that correlations (mostly between $\alpha(E)$, $\eta(E)$ and $g$), introduced once a certain operational gain is required, invert the explicit dependency with the gap size given by eq. \ref{TimeRPC}: experimentally, the best time resolutions are found for narrow gaps.\footnote{For a simple image note that the product of $\alpha(E)-\eta(E)$ and $g$ is bound to an approximately constant value, that can be intuitively viewed as some form of `Raether criteria'.} 

It is strongly hinted by data and models \cite{DiegoPro, JB, Repond} that the main source of degradation of timing RPCs at high particle flux
is due to the average field drop caused by the current flowing through the resistive plates, with event-by-event fluctuations playing a minor role:
\beq
\overline{\Delta{E}} = \frac{1}{g} \bar{q} \bar{\phi} \rho d^* \label{Drop}
\eeq
Here $\bar{q}$ is the average charge per gap, $\rho$ the resistivity of the plates and $d^*$ their equivalent thickness per gap, with $\bar{\phi}$ being the average particle flux over the irradiated area \cite{JB}:
\beq
\bar{\phi} = \frac{\int\int (\phi(x,y))^2 dx dy}{\int \phi(x,y) dx dy}
\eeq
The above approximation is often referred to as `the DC model'. By assuming a recursive dependence of $\bar{q}$ with the field, $\overline{\Delta{E}}$ can be exactly obtained in a number of practical cases \cite{Laso, DiegoPro, Benciveni}, but in the following argument we consider the case of small field drops, where $\bar{q}$ can be taken as approximately constant.
After introducing the most recent low-pressure measurements for the main RPC gas (C$_2$H$_2$F$_4$) of $\alpha(E)-\eta(E)$ and $v_d(E)$ \cite{Urquijo}, we come to:
\beq
S(E) = (\alpha(E)-\eta(E))v_d(E) \simeq \frac{dS}{dE}(E-E_{th})
\eeq
that is valid in the typical operating regime of timing RPCs ($E\simeq100$ kV/cm). The following expressions then follow:
\bear
\frac{\Delta\sigma_t}{\sigma_t}(\bar{\phi}) \simeq && \frac{1}{S_o}\frac{dS}{dE}\overline{\Delta{E}} \\
                                              \simeq && 50\% \times \frac{\bar{\phi}}{1 ~ \tn{kHz/cm}^2} \times \frac{\rho d}{3 \cdot 10^{12} ~ \Omega \tn{cm} \times \tn{mm}} \label{SigmaFlux}
\eear
For the evaluation of the above expression we have taken typical values: for the charge $\bar{q}=2$ pC, for the ionization rate $S_o$ we assume a typical field of $100$ kV/cm, and finally $g=0.3$ mm.
The column resistivity appears referred to that of 1 mm of float glass ($3 \cdot 10^{12}~ \Omega \tn{cm} \times \tn{mm}$).

Through expression \ref{SigmaFlux} it is hence made explicit that operation at 1~kHz/cm$^2$ can be only reasonably achieved for very thin float-glass, clearly being at the limit of the technology in any case. Chinese glass, with a value for $\rho = 3 \cdot 10^{10} ~ \Omega \tn{cm}$ (at 20$^\circ$C) can improve this figure by approximately 2 orders of magnitude. Materials with resistivities below $\rho \simeq 10^{10} ~\Omega \tn{cm}$ often show poor streamer quenching, resulting in a worsening of the time resolution. Therefore, with today's understanding, a maximum particle flux around 100 kHz/cm$^2$ represents a good estimate of the RPC technological horizon. A version of the scaling given by eq. \ref{SigmaFlux} for a number of materials can be found in \cite{CBM}.

\subsection{MPGDs}\label{MPGDs}

MPGDs make extensive use of photo-lithographic techniques in their fabrication, but the range of construction techniques has also expanded towards providing resistive protection either via embedded resistors, screen-printing, surface dissipation over thin layers, or a combination of them (see for instance \cite{Max} and references therein). In some configurations (e.g., triple-GEM), discharge probability can be made negligible small for gains around 10$^5$ even in the presence of highly ionizing radiation \cite{CMS}. The ultimate rate capability of MPGDs when globally considering the effect of either a resistive protection or the damage caused by residual discharges is difficult to quantify at present, but results recently obtained under $5.9$ keV X-rays in \cite{Fil} ($\sim250$ initial electrons) indicate that they can reliably operate up to at least 10-100~MHz/cm$^2$ when irradiated by mips. Apparently, the only insurmountable problem for accomplishing timing in these harsh conditions is the physical jitter on the collection time of the ionization produced in the so-called `drift region'. A solution to this long-standing problem, that had already shown a time resolution of 680ps per single electron in early detectors \cite{Charpak1}, has been recently re-introduced by making use of modern manufacturing techniques.

The idea here is to resort to the emission of Cherenkov photons when the impinging particle crosses an auxiliary MgF window, photons that are converted to electrons via a CsI layer in contact with the gaseous medium. The proposal embodies two additional regions (`drift' and `amplification') separated via meshes in a Micromegas-fashion, a geometry aimed at providing the necessary flexibility in terms of ion and photon-feedback suppression. In a recent work \cite{ThomasLast} a time resolution per single photo-electron of 200 ps was achieved, a value largely determined, according to simulation, by the diffusion of the electrons during their transit along the drift region:
\beq
  \sigma_t(E) = \frac{A}{\sqrt{n_{pe}}} \frac{D_L^*(E)}{v_d(E)} \sqrt{g} \label{ResMM}
\eeq
In eq. \ref{ResMM} $n_{pe}$ is the average number of photo-electrons, $D_L^*$ the longitudinal diffusion coefficient in units of $[L^{1/2}]$, $v_d$ the electron drift velocity and $g$ the thickness of the drift region \cite{ThomasLast}. According to simulation, the presence of pre-amplification improves the time resolution by a value of typically $A\sim0.5$-$1$. The good statistical scaling with $\sqrt{n_{pe}}$ projects a time resolution at the scale of 20-40 ps for minimum ionizing particles when using windows with thicknesses at the few mm-scale, something that is currently being studied in beam.

\section{Timing on large areas}\label{LargeArea}

Once a given position resolution is targeted by design, and as long as particle multiplicity is low, strip layouts show a linear scaling with the detector size and pads a quadratic one. Besides, specially for a timing device, there is a natural limit to the maximum area of the pad due to capacitive effects, that does not exist in the case of strips, or at least it manifests itself much more mildly \cite{DiegoCurr}. Hence, and by virtue of their large simplicity of construction, we focus here on the `multi-strip RPC' concept \cite{FOPIandreas}.

\subsection{Modal dispersion and its compensation}

\begin{figure}[h!]
\centering
\includegraphics*[width=5.3cm]{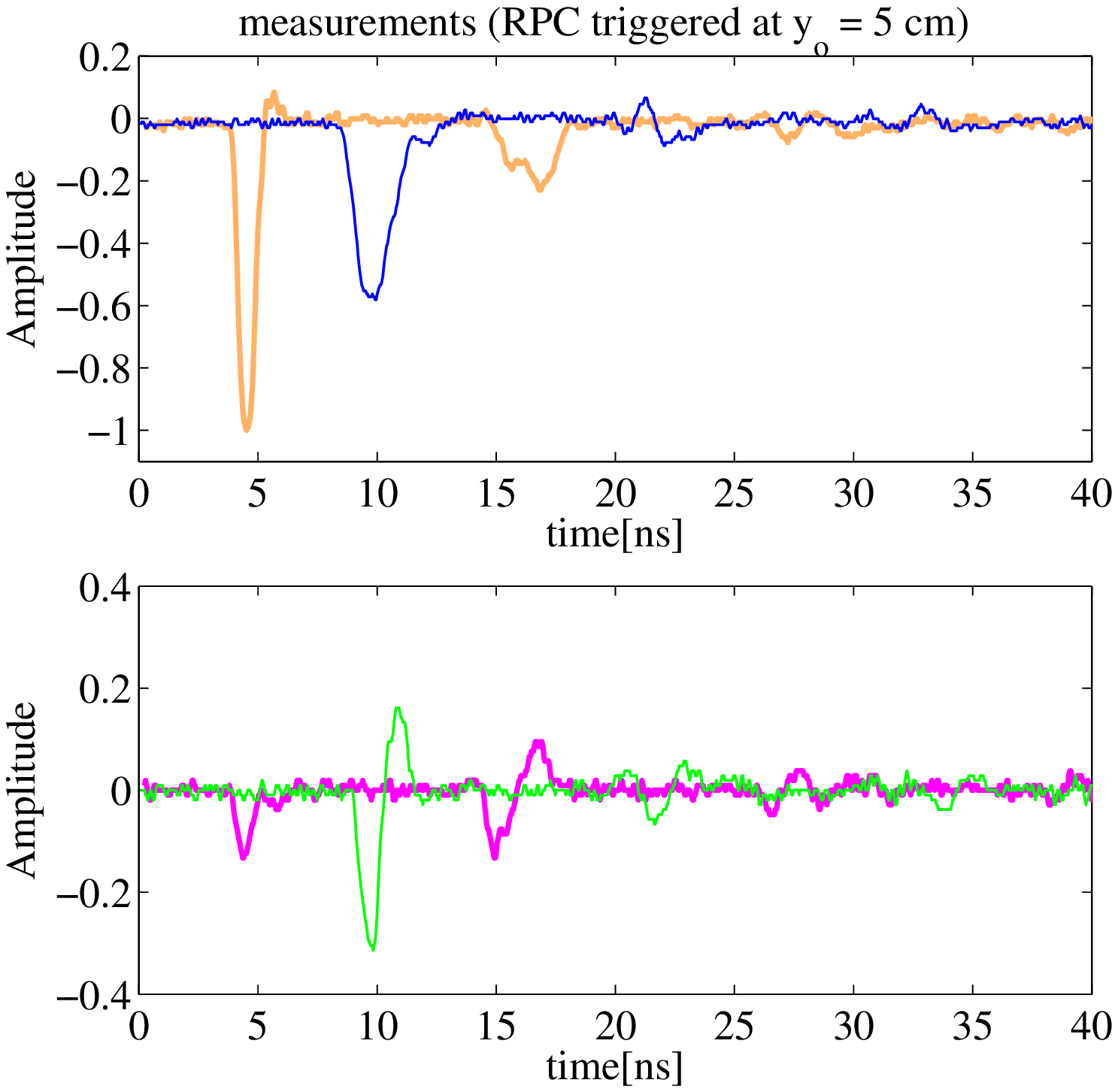}
\includegraphics*[width=5.3cm]{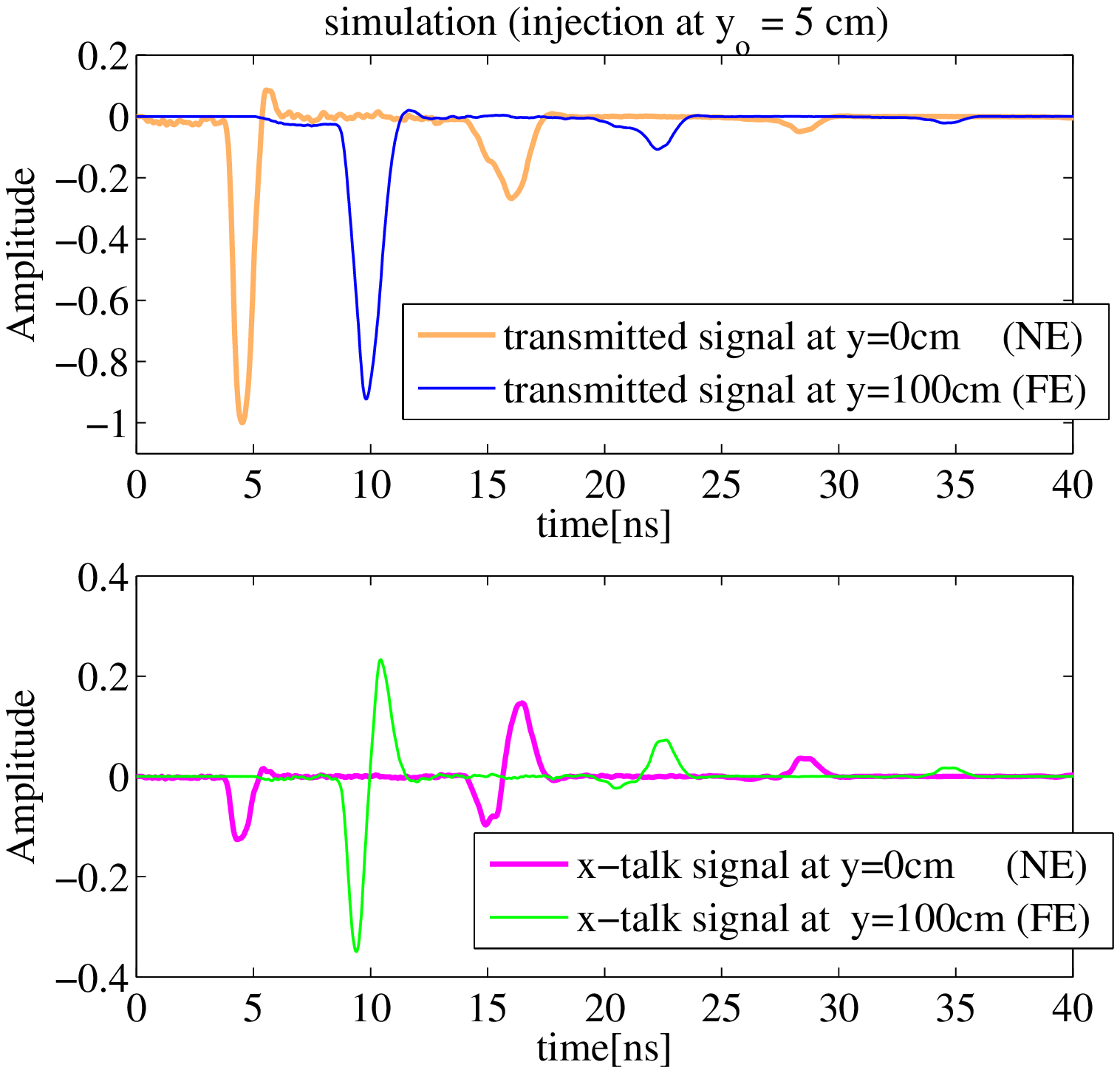}

\includegraphics*[width=5.3cm]{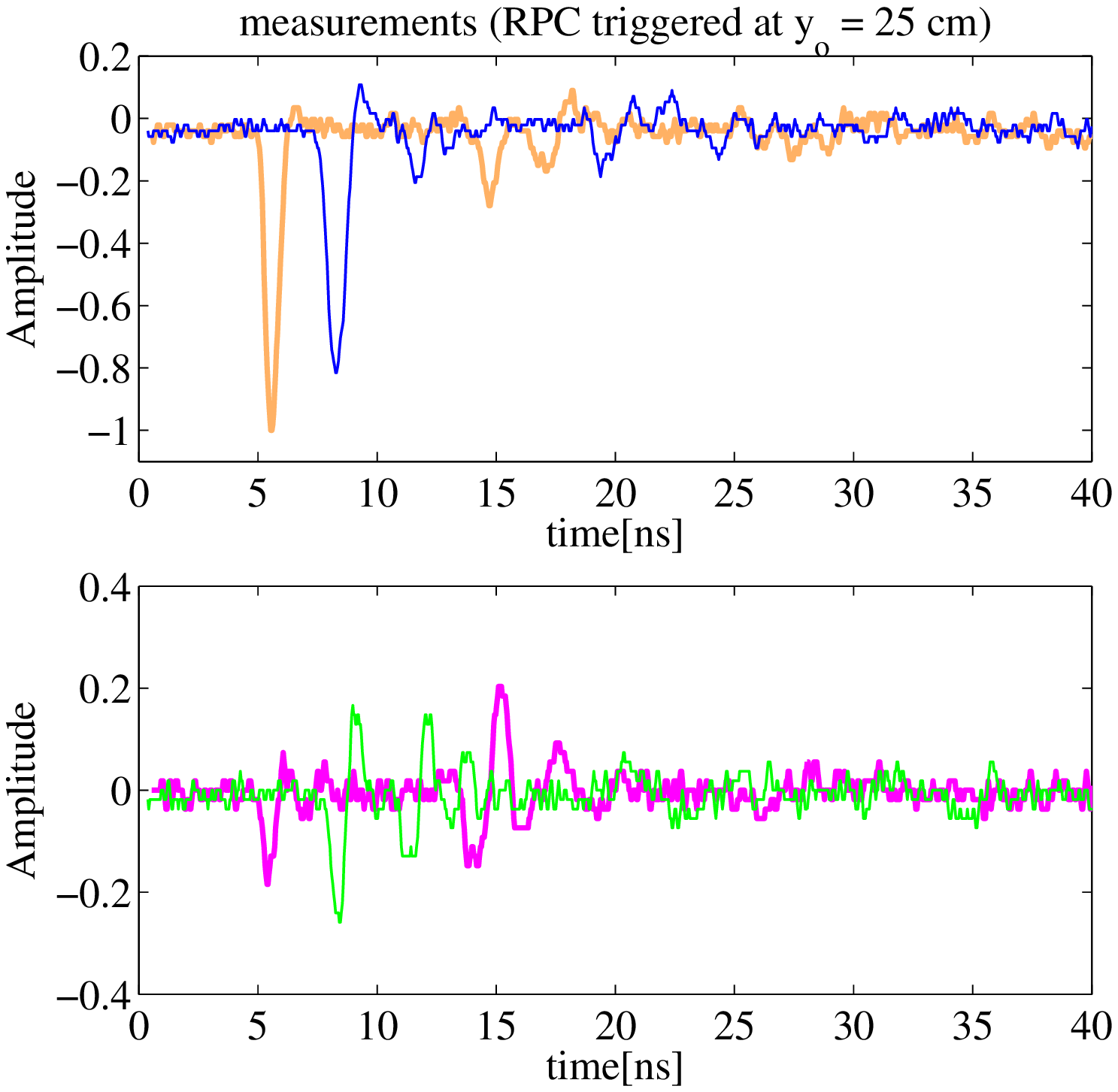}
\includegraphics*[width=5.3cm]{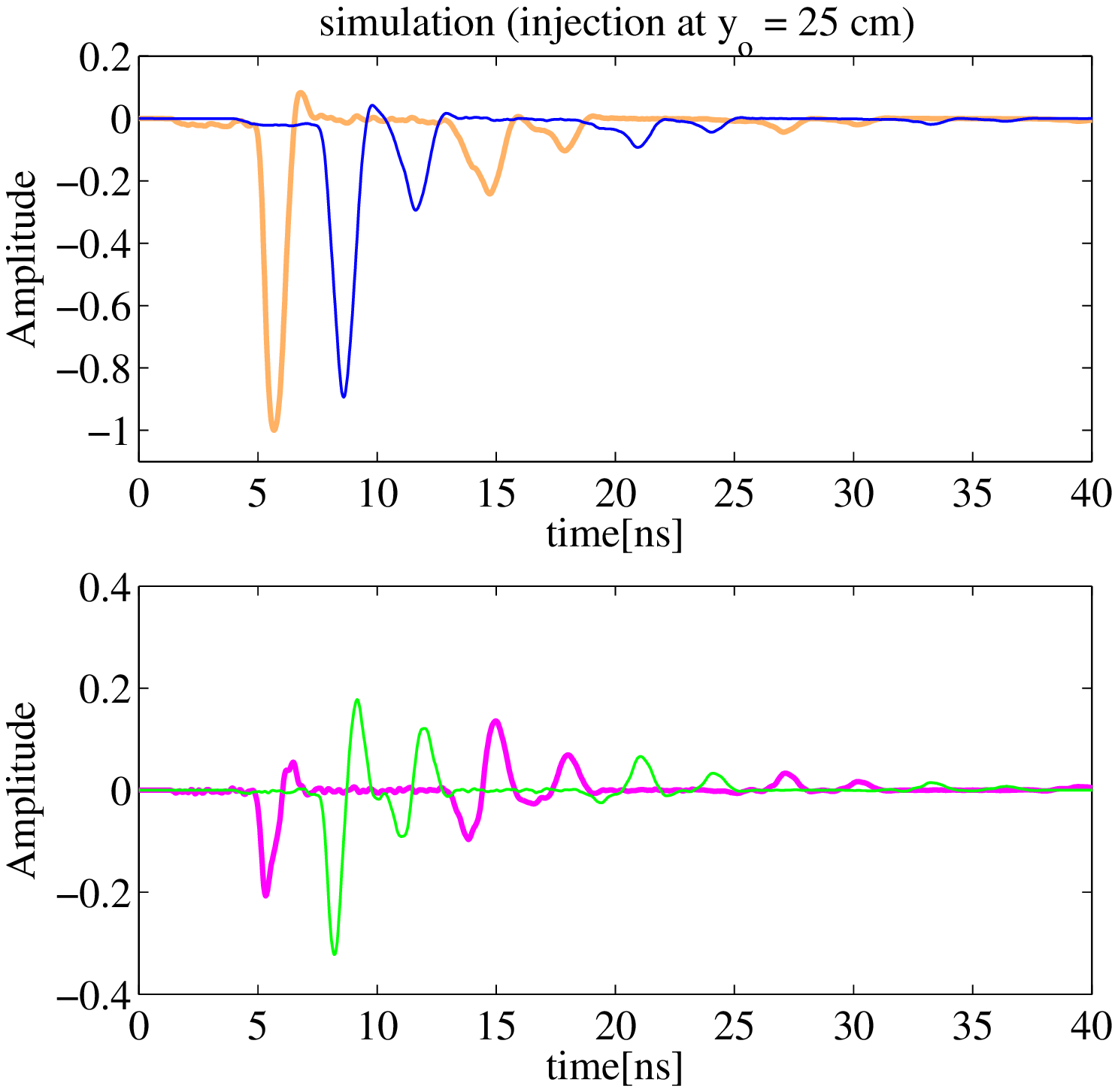}

\includegraphics*[width=5.3cm]{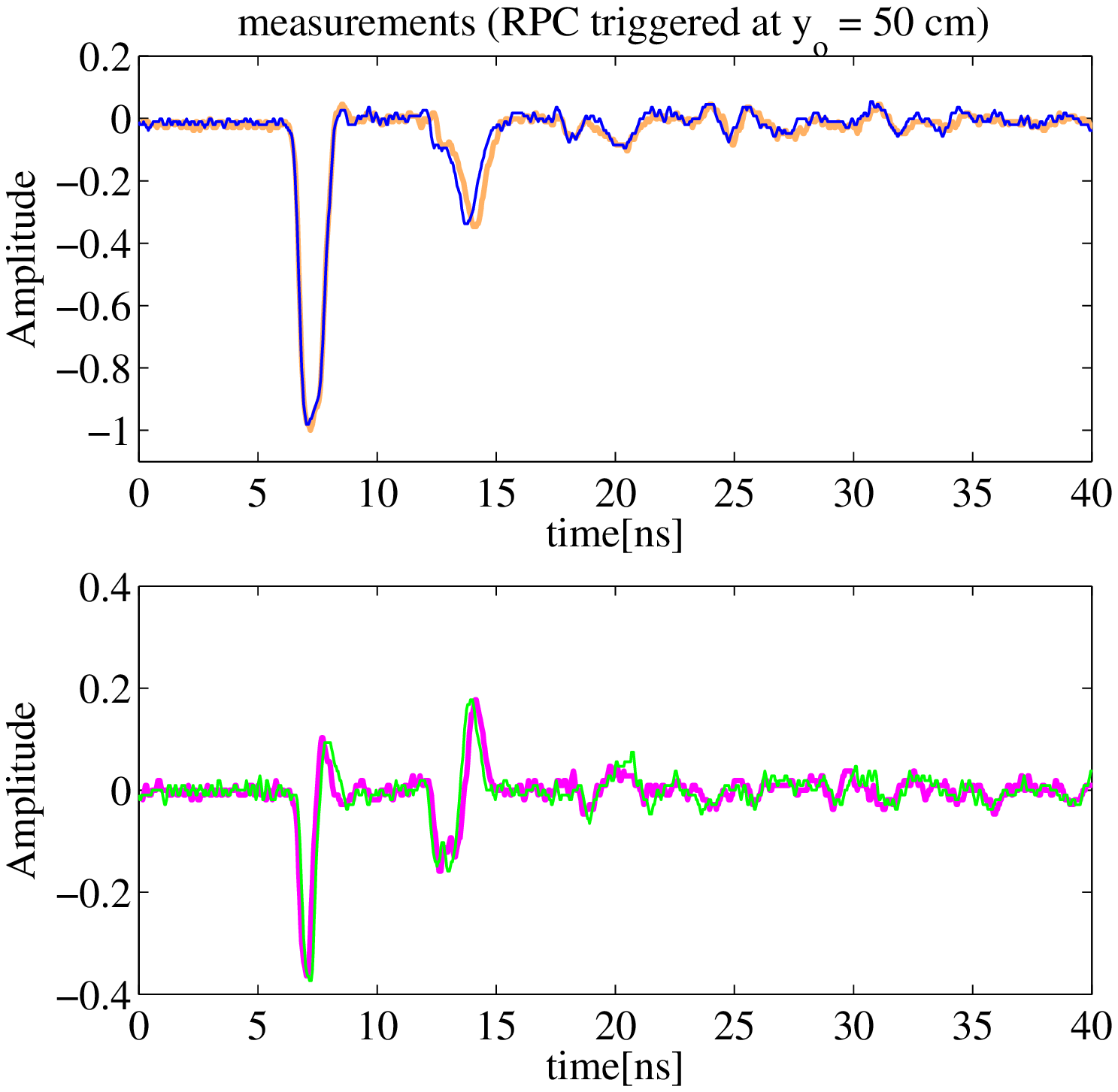}
\includegraphics*[width=5.3cm]{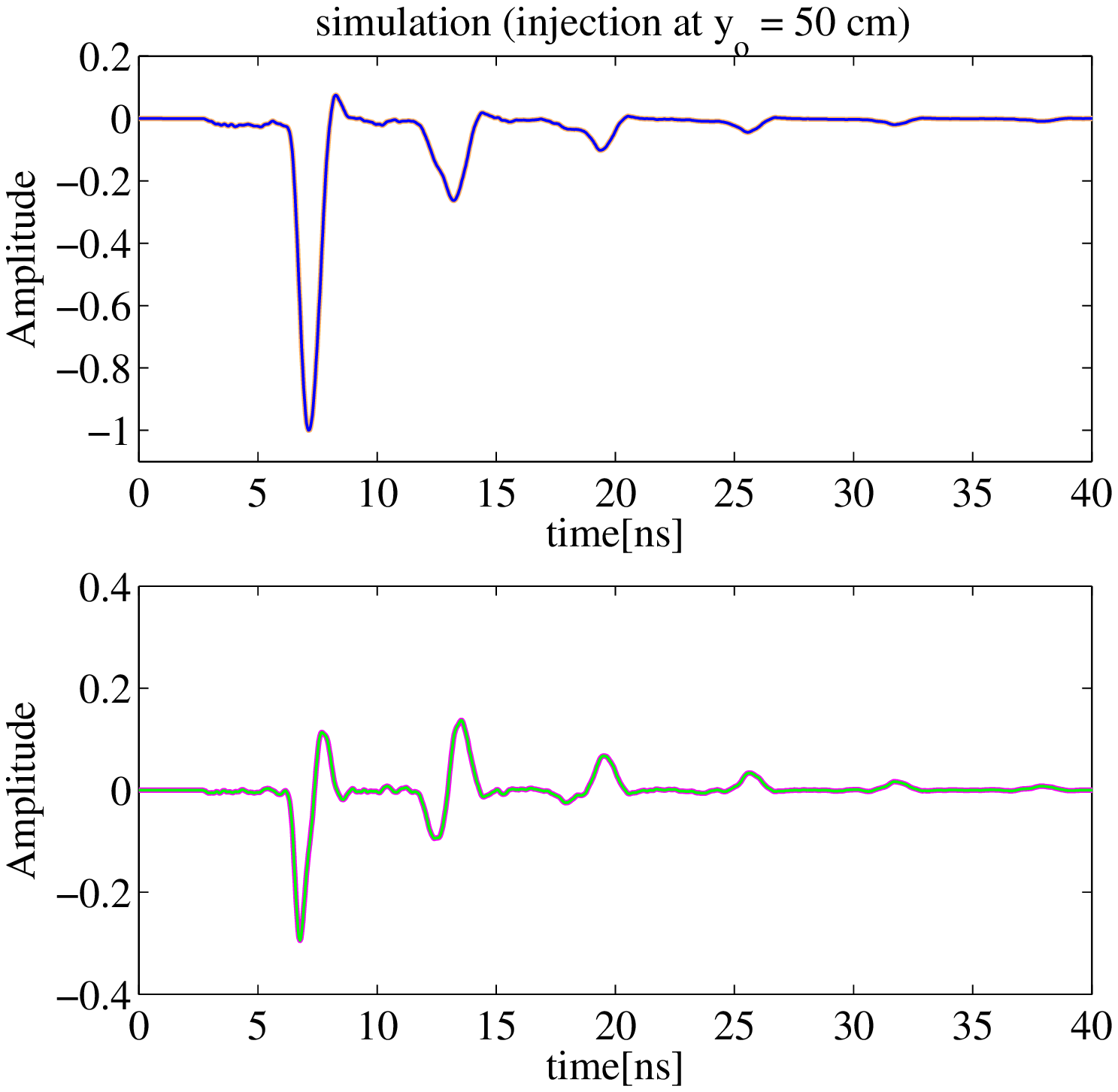}
\caption{Left: signals measured in the 1m-long BGOegg prototype for different trigger positions ($Area=1$cm $\times$ 1cm, and centered relative to a given 2.5cm-wide reference strip), with a cosmic ray stand \cite{Natsuki}.
The signals are measured at the near-end (NE) and far-end (FE) of the triggered strip (orange/blue) and the first neighbour (pink/green), for three characteristic positions relative to the near-end (5cm, 25cm, 50cm). Right: simulation performed using the same technique as in \cite{DiegoIntegrity}, with dielectric parameters: $\tan{\delta}=0.025$, $\epsilon_r=5.5$ for float glass, $\epsilon_r=4.4$ for FR4 \cite{DiegoIntegrity} and $\epsilon_r=10$ for the HV graphite tape. Each series is normalized to the amplitude of the signals transmitted at the near-end (orange) and, of those, the one corresponding to $1$cm (measured with minimal shaping and not shown here) is used as input for all simulations. Most of the discrepancies seem attributable to the lack of a proper description of the tapered end of the strip (that influences the far-end of both the transmitted and cross-talk signals in the 5cm case, due to an imperfect description of the superposition between direct and reflected waves), the absence of an unbiased estimate of the input current, the small offsets in cables, the position jitter within the trigger region and white noise.}
\label{NatsukiFig}
\end{figure}

Multi-strip RPCs are highly dispersive devices in a form commonly known as `modal dispersion' \cite{Clayton}, a phenomenon that can severely impair operation. Apparently, the notion of the possible existence of modal dispersion in gaseous detectors was first indicated by W. Riegler in \cite{Riegler}, indirectly confirmed in \cite{DiegoFirst} and finally measured in \cite{DiegoIntegrity} on a 2m-long double-strip timing RPC excited externally. During the XIII RPC Conference several groups reported high levels of cross-talk and a strong dependence of the time resolution with the position of the impinging particle \cite{NatsukiProp, NatsukiProp2}, a text-book example of the manifestation of modal dispersion. Transmission and cross-talk patterns observed and simulated in one of these detectors (BGOegg) are given in Fig. \ref{NatsukiFig} for signals induced at three positions along the strip.

\begin{figure}[h]
\centering
\includegraphics*[width=12cm]{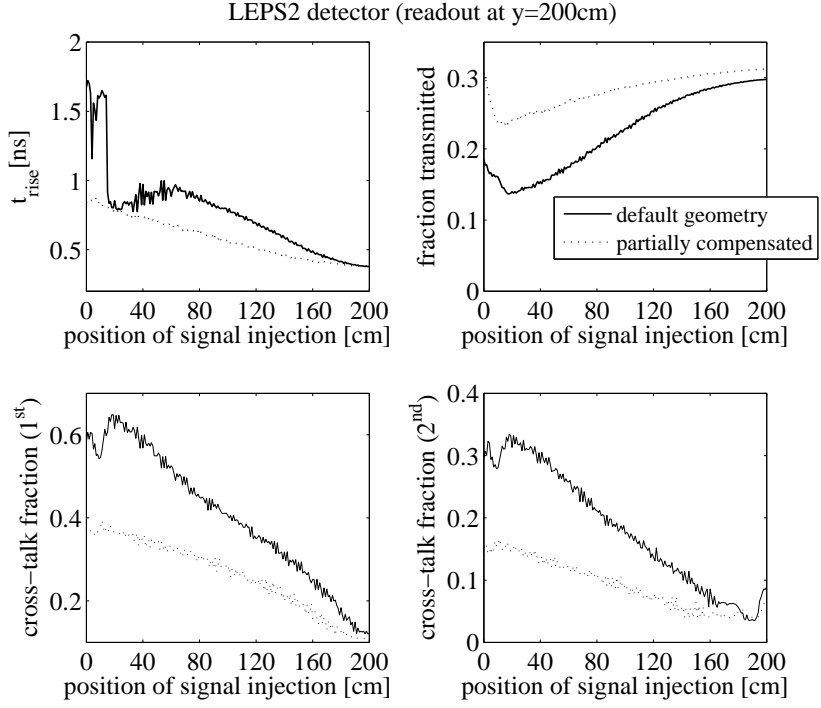}
\caption{Main signal characteristics as a function of the propagation distance for the LEPS2 detector. From top-left to bottom-right: signal rise-time (from 10\% to 90\% of the signal maximum), fraction of transmitted signal, fraction of cross-talk signal in $1^{st}$ neighbour relative to transmitted signal, the same but for the $2^{nd}$ neighbour. The continuous line shows the result for the default geometry, and the dashed line the expected behaviour after a simple structural modification that would reduce the strength of modal dispersion by about a factor $\times 2$.}
\label{NatsukiFig2}
\end{figure}

A priori, modal dispersion is a feature inherent to the propagation along any inhomogeneous dielectric that can be assimilated to a multi-conductor transmission line (in particular, a multi-strip RPC). It results from an imbalance between the capacitive and inductive coupling of the line, a condition that reads, with approximate character, as:
\beq
\frac{C_m}{C_0} \neq \frac{L_m}{L_0} ~~~~~~~~~~~~~~~~~~(\tn{and it is useful to recall that:} ~~~~ \frac{L_m}{L_0} \simeq \frac{C_m}{C_0}\bigg|_0) \label{CompCondition}
\eeq
In eq. \ref{CompCondition} $C_m$ refers to the mutual capacitance (capacitance between neighbouring strips), $C_0$ is the total strip capacitance, $L_m$ is the mutual inductance between neighbouring strips and $L_0$ is the strip self-inductance; an additional sub-index $0$ refers to the values obtained by re-placing all dielectrics by vacuum. Under approximations of general validity for RPCs (`short-range coupling': $\frac{C_{m,2}}{C_m}, \frac{C_{m,2}}{C_m}\big|_0\ll1$),\footnote{$C_{m,2}$ refers to the capacitance between a strip and its second neighbor.} it can be reasoned, without any further approximation \cite{DiegoIntegrity}, that the signal breaks during propagation on as many modes as transmission strips ($N_s$), travelling with velocities ($\vec{v}$) that are functions of the imbalance given by eq. \ref{CompCondition}:
\bear
&& \frac{\vec{v}}{c} = \sqrt{\frac{C_{00}}{C_0}} \vec{f}(\frac{\Delta{v}}{\overline{v}}) \\
&& \frac{\Delta{v}}{\overline{v}} = \frac{C_m}{C_0} - \frac{C_m}{C_0}\bigg|_0
\eear
with $\vec{f}$ having dimension $N_s$ and satisfying $\vec{f}(0) = \{1,...,1\}$, and $c$ is the speed of light. The magnitude $\Delta{v}/\overline{v}$ takes a precise meaning in the 2-strip case, where it represents the relative velocity spread between the two system modes with $\overline{v}$ being their average velocity \cite{DiegoFirst}. For a generic $N_s$-strip RPC one can still make use of this magnitude to define a characteristic delay per unit length as:
\beq
\Delta{T_y} = \frac{1}{\overline{v}} \frac{\Delta{v}}{\overline{v}}
\eeq
As an example, $\Delta{T_y}$ is at the level of 3.5~ps/cm in the geometry studied in Fig. \ref{NatsukiFig}. The net effect of this delay between the modes is the increase of the signal rise-time with the propagation distance $y_o$ (Figs. \ref{NatsukiFig}, \ref{NatsukiFig2}). It also plausibly explains the worsening of the time resolution as a function of the distance to the strip ends in 2m-long counters \cite{NatsukiProp}. This systematic dependence shifts the time resolution from an intrinsic value around 60~ps to slightly above 80~ps at the counter ends, thus compromising the goals of the LEPS2 experiment, for instance. Moreover, in simulation it can be seen that a critical condition exists when
$y_o \Delta{T_y} \gtrsim t_{rise}$, implying that the modes begin to fully decouple. This change in behaviour can be clearly seen in Fig. \ref{NatsukiFig2} for propagation distances beyond 140 cm.

What is important to note is that $\Delta{v}/\overline{v}$ can be made zero not only in the theoretical limit where the transmission structure is a homogeneous dielectric: one finds indeed a number of possible realizations of this condition in a practical detector. We refer to the restoration of this ideal situation as `compensation'. Compensation in timing RPCs was demonstrated in \cite{DiegoIntegrity}, and later employed in the STAR-MTD detector \cite{STARHS}. In the LEPS2 case (Fig. \ref{NatsukiFig2}), it is relatively simple to modify the detector geometry by interleaving a thin teflon sheet around the central electrodes, thus decreasing the modal dispersion to about 2~ps/cm, and enabling again a reasonable transmission over 2m.

The difficulty at conceiving that, in reality, a signal can possibly travel as a decomposition of as many modes as strips (specially being the furthest strip maybe up to one meter apart), was already addressed in one of the first papers on this topic \cite{DiegoIntegrity}. Basically, in extreme scenarios, and due to the existence of a finite propagation speed, an electrical perturbation may only know about the existence of the furthest strip once it is already exiting the structure!. Because Transmission Line Theory assumes electrostatics at every position during propagation, it is implicit to its use that the structure must be electrically short on the transverse dimension, a condition that implies for timing RPC signals a distance of about $\Lambda_e=5$cm \cite{DiegoIntegrity}. This led some authors to argue about the possibility of having `not fully organized modes' \cite{FonteProp}. It is important, however, to realize that by virtue of the fact that most RPCs fall under the `short range coupling' approximation ($\frac{C_{m,2}}{C_m}, \frac{C_{m,2}}{C_m}\big|_0\ll1$), the most important coupling takes place to the nearby strips, and those are in any typical detector within the aforementioned distance. Although this explanation may seem satisfactory, and qualitatively explains the good agreement observed between theory and measurements in the 8-strip detector simulated in Fig. \ref{NatsukiFig}, it still does not provide an answer to the remarkable fact that the travelling signals are decomposed precisely in 8 modes. Only a full 3D-simulation will further clarify these aspects.

\subsection{A technological `twist': the Serpentine (Delay-Line) -RPC}

Delay-line electrodes represent a relatively popular readout configuration aimed at reducing the number of electronic channels. The geometry is not completely alien to the RPC field, where it is employed to cover large areas for efficiency studies \cite{ItalianLady}. Without additional modifications, however, RPC detectors are not suited to build well-behaved delay lines, due to the dispersive character introduced by the modal dispersion and its dramatic effects on the inter-strip cross-talk. A simple derivation of the fraction of cross-talk expected for exponential signals ($I \sim \exp[(\alpha-\eta)v_d t]$) is given in \cite{DiegoFirst}:
\bear
\hat{F}_{ct} = \frac{\hat{I}_{ct}}{\hat{I}_{tr}} \simeq \frac{R}{Z_c + R} \frac{C_m}{C_0} + \frac{1}{2} (\alpha-\eta)v_d \times y_o \Delta{T_y}
\eear
where $R$ is the amplifier's input resistance and $Z_c$ the characteristic impedance of the line. Here, again, we see a dispersive term proportional to the quantity $y_o \Delta{T_y}$ and to the signal rise-time. If this term can be made zero while the first term (capacitive cross-talk) is reduced to a \%-level, a well-behaved  Serpentine-RPC would seem realizable.\footnote{Here we are neglecting charge sharing between strips, an effect that is small overall for wide strips, but escapes nonetheless from the scope of the present work.} Signal
transmission under these conditions will be mainly affected by losses and by the reflections introduced by the bends. Losses during transmission show cutoff frequencies (defined as a 3dB signal drop) due to resistive (sub-index $R$) and dielectric (sub-index $G$) behaviours, and can be approximated by \cite{DiegoIntegrity}:
\bear
&&f_{c,R} \simeq \left(\frac{Z_c \tn{w} \ln{2}}{D} \right)^2 \frac{2}{\pi\mu\rho_{_{DC}}} \label{fc_R}\\
&&f_{c,G} \simeq \frac{\bar{v}\ln{2}}{2\pi D \tan{\delta}^*}  \label{fc_G}
\eear
Here $w$ is the strip width, $D$ the detector length, $\rho_{_{DC}}$ the DC resistance of the strips and $\mu$ their magnetic permeability. The dielectric losses are determined by the effective $\tan{\delta}^*$ of the structure, around 0.01 at 1GHz for float glass-based RPCs. By inserting standard values, it can be seen that the bandwidth of the structure (referring here to end-to-end propagation) is not smaller than 200 MHz for an overall length of 10 m,
a bandwidth that has been shown to be compatible with 50ps-timing (e.g.: NINO, PADI \cite{NINO, PADI}). This time jitter should be sufficient to reconstruct the crossing point of a traversing particle with a 5 mm precision.

\begin{figure}[h]
\centering

\includegraphics*[width=7.5cm]{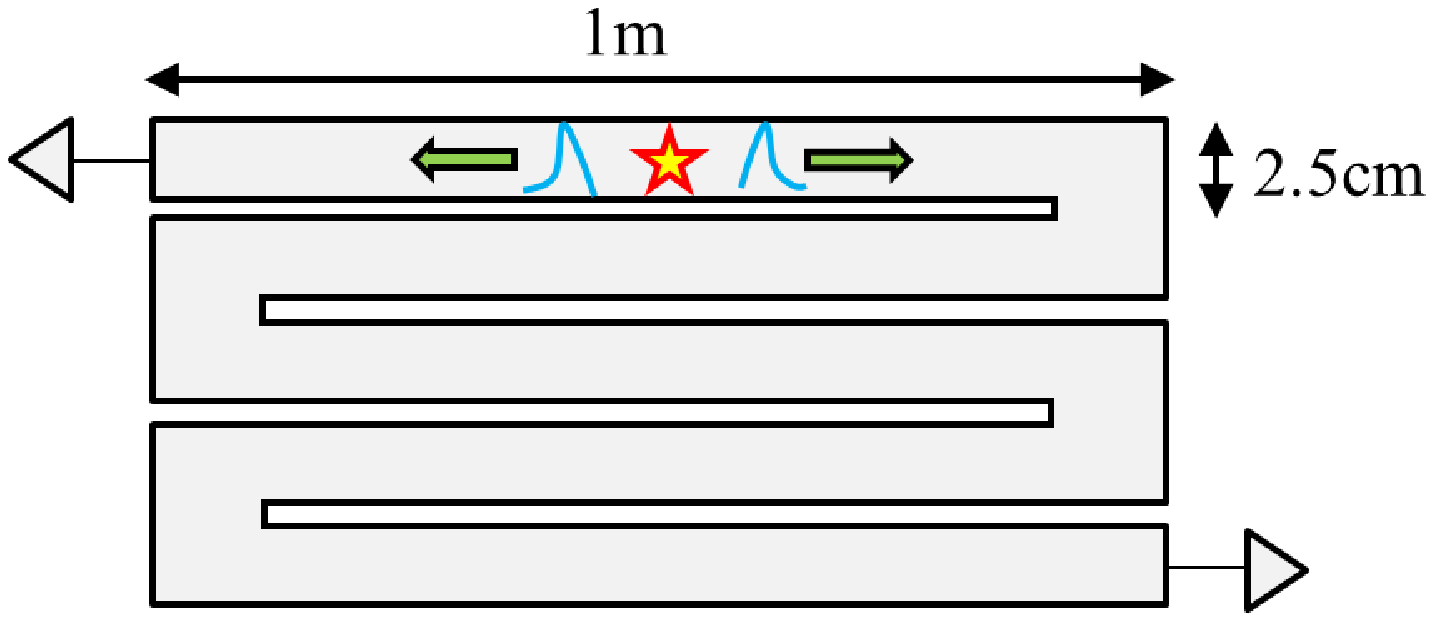}
\includegraphics*[width=7.5cm]{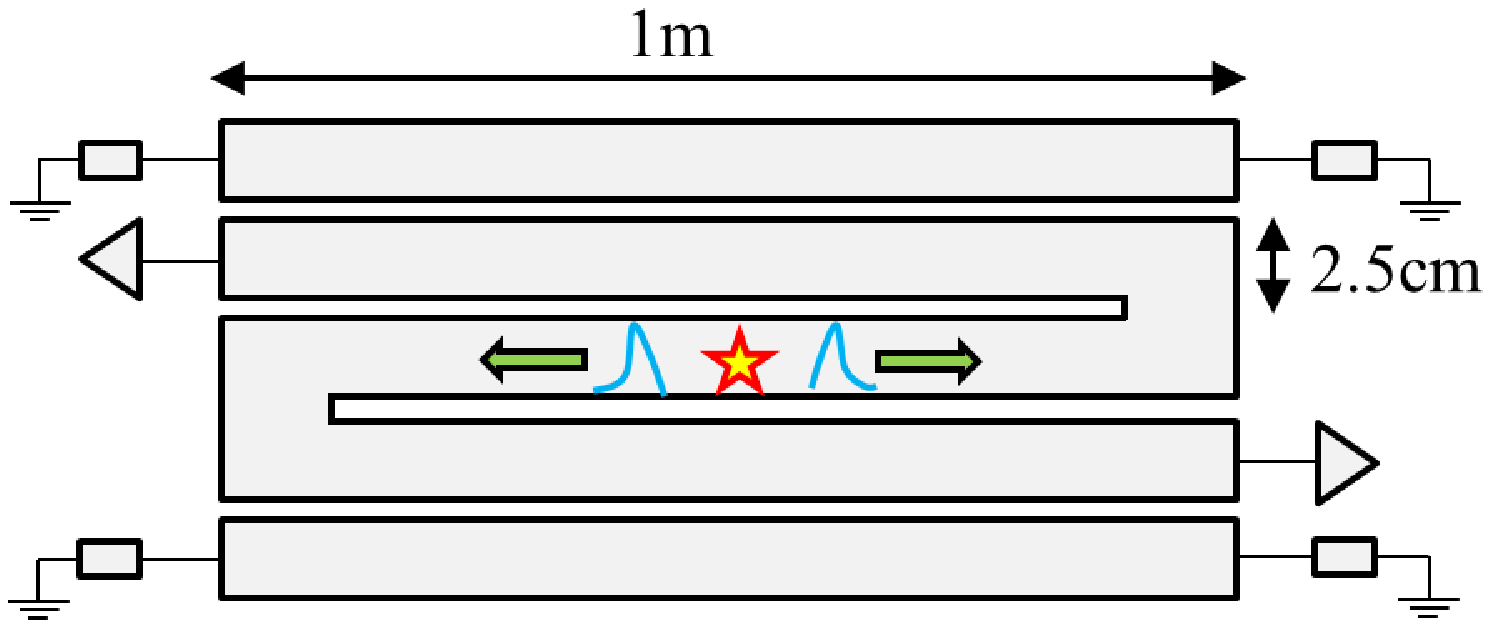}

\includegraphics*[width=7.5cm]{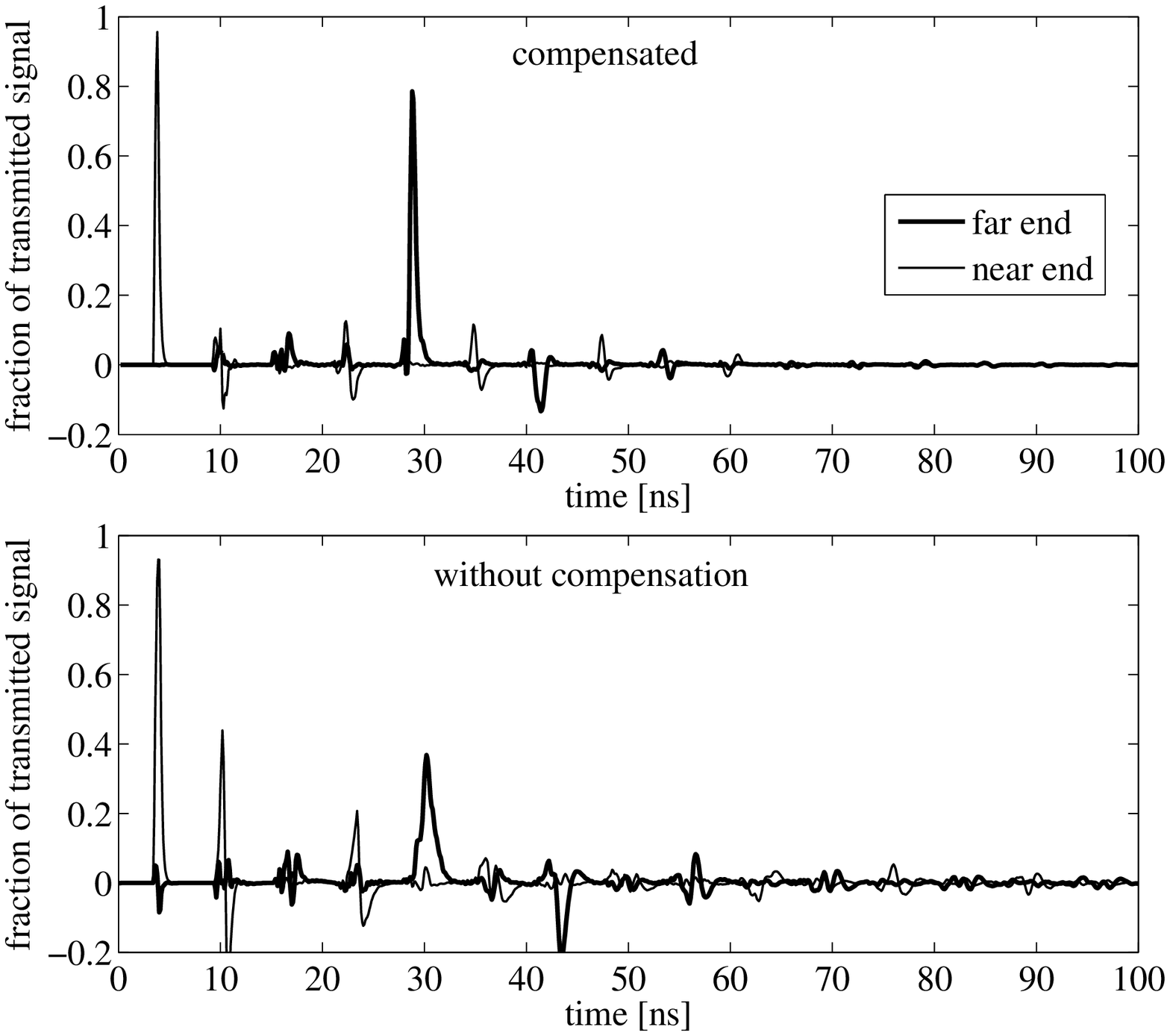}
\includegraphics*[width=7.5cm]{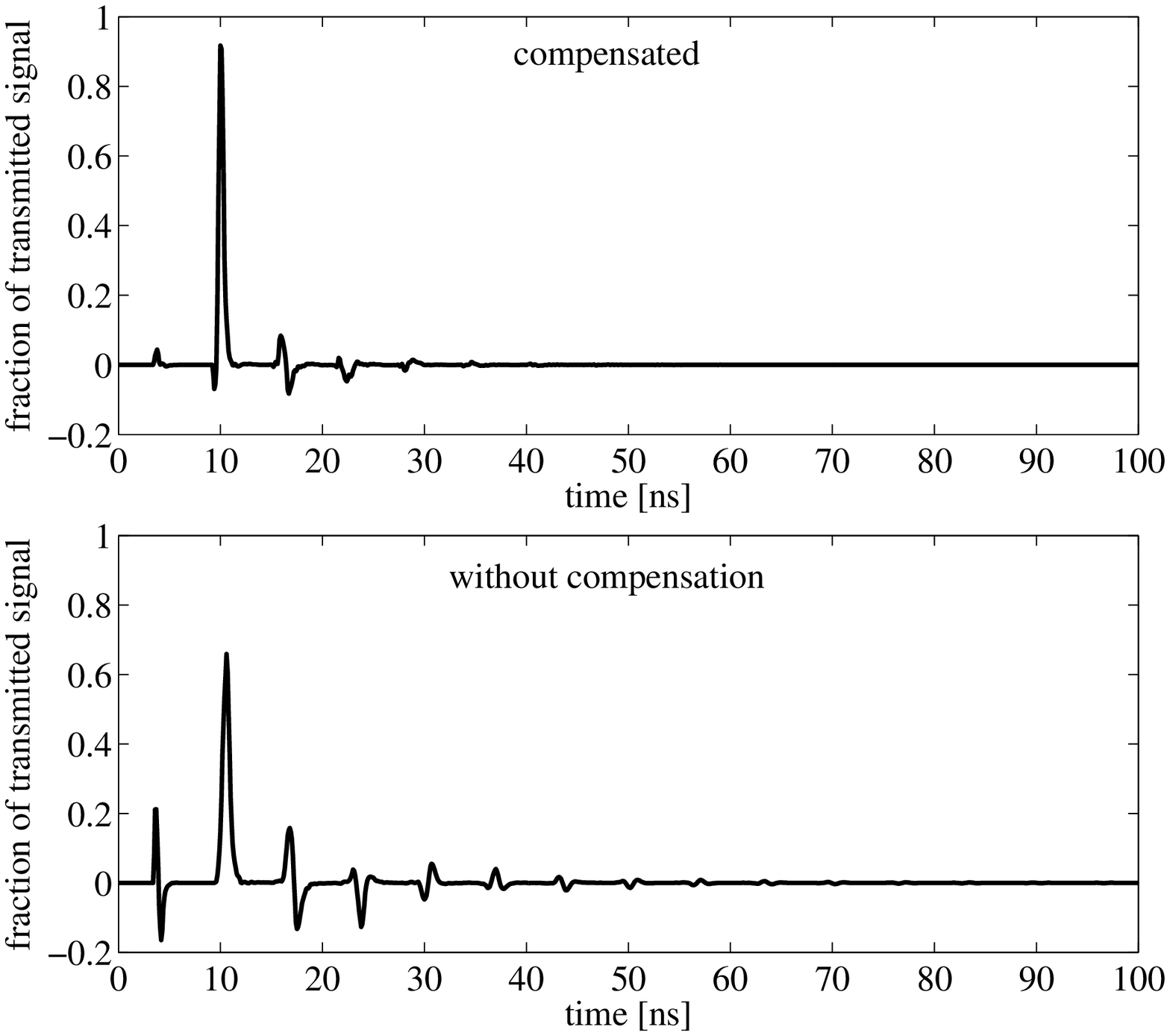}

\caption{A 5-strip Serpentine-RPC (left), with and without compensation. Signals are injected as indicated in the drawings and read at the near and far ends. On the right a 3-strip Serpentine-RPC with additional guard-strips is shown for comparison. The presence of spurious pulses before the arrival of the interesting signal is below 5\%. No optimization of the coupling coefficient $C_m/C_0$ or the shape of the bends has been performed.}
\label{Rogelio}
\end{figure}

Based on these arguments, it seems most natural to consider the possibility of a 10-fold reduction of the number of channels in a 1m-scale detector by means of the serpentine approach, although the concept allows easily a 100-fold reduction at the 10cm-scale, and so on.
To illustrate that, we performed a simulation based on the Allegro suite with its FEM solver EMS2D \cite{Allegro}. The simulations were benchmarked against earlier simulations performed with MAXWELL-2D and a custom MTL-solver in \cite{DiegoIntegrity}, yielding comparable results. The bend-model used is described in \cite{BendModel}. We studied the compensated structures given in \cite{DiegoIntegrity}, in particular the 4-gap strip-line RPC
described in Fig. 14b of that work. The number of possible implementations of a delay line is very broad, so here we give only two exemplary cases:
Fig. \ref{Rogelio} shows the transmission over a 5-strip (left) and 3-strip (right) Serpentine-RPC for a typical RPC signal as seen with a bandwidth of 1GHz. For the 5-strip structure injection is performed at the middle of the upper strip, and for the 3-strip one at the middle of the structure. All signals
are normalized to the injected one and the line is matched through the amplifier $R=Z_c$. The effect is
very clear: in the absence of compensation the signal dispersion is very strong, with spurious pulses appearing in the range of 20-40\% of the signal amplitude, and this is reduced to 10\% after compensation. A better job can be performed in the presence of `guard strips' (those could be added outside the active region or could
be part of another nearby serpentine). In this latter case the spurious pulses arriving prior to the signal are below 5\% (right figure).

These results are just representative of
what can be done if a compensated Serpentine-RPC would be further optimized to reduce the capacitive coupling $C_m/C_0$ and the effects of the bends.
These type of optimized structures are currently being designed with the aim at its experimental validation in the context of muon tomography and muography.

\section{The technological horizon}\label{Horizon}

Based on simulations, and assuming that all aforementioned technical caveats can be solved, the theoretical technological limit of an approach based on present MPGD technology may be put at a factor $\times 1/4$ below the demonstrated value of $200$ps-$\sigma$/photo-electron. A factor $\times 1/2$ can be gained if the drift region can be reduced from $200\mu$m to $50\mu$m, and another factor $\times 1/2$ if operation under pure quencher is feasible \cite{ThomasLast}. The final performance for minimum ionizing particles will depend on the amount of photons produced at the radiator crystal and the photo-effect quantum efficiency, potentially reaching $n_{pe}=40$ for a good photocathode. Being the device limited mainly by diffusion and showing a good $1/\sqrt{n_{pe}}$ scaling, the projections for the time resolution seem robust. Importantly, however, the necessary sensitivity and both the short and long-term stability need to be demonstrated.

Concerning RPCs, where experimental information abounds, the best present value for the time resolution is 21ps-$\sigma$ for minimum ionizing particles, achieved in \cite{Williams20} on an ensemble of 24 gas gaps of $160\mu$m size. Values around 50ps represent already a technology standard \cite{JB, Babkin}. Narrower gaps have been tried, but never down to the ${50\mu}$m or even $25\mu$m values that are characteristic of $\mu$-bulk Micromegas, for instance \cite{PacoMicroBulk}. $\mu$-bulk Micromegas detectors show an excellent mechanical accuracy (\cite{Elisa}) and nothing seems to impede at the moment the use of similar structures or materials as RPC-spacers themselves. It will suffice to have 24 gaps of ${50\mu}$m size for reaching a total gas column of 1.2mm, considered to be the minimum for efficiency to mips (equiv.: $0.3$mm $\times 4$ gaps). It is at the moment not possible to anticipate the achievable improvement, and to which extent the necessary operation under extreme space charge conditions could limit performance \cite{WernerChristian}, but it certainly deserves exploration. Narrower gaps are expected to yield faster signals, and therefore the necessity of electrostatic compensation ($C_m/C_0=C_{m0}/C_{00}$) will turn essential for m$^2$-counters.

Another possible area of exploration, long neglected since the first Pestov counters, is the possibility of operation at high pressure. Higher pressure can lessen the number of gaps needed for efficiency, and can be generally considered to be an advantage, according to eq. \ref{TimeRPC}. The original (single $100~{\mu}$m-gap) Pestov counters operated at 12 bar, and while this extreme number was then necessary for an efficient detector, the situation is much more comfortable with the multi-gap technology. Operation at 2-4 bar would thus enable the use of ${25\mu}$m-${50\mu}$m gaps, based on conventional 4-10 gap architectures that are well established at atmospheric pressure. On the other hand, it has been recently shown that operation up to 4 bar in homogeneous fields does not show indications of a strong reduction of the working gain, both for Xenon and Argon-based mixtures in the case of micromegas with $\sim50$-$100\mu$m gaps \cite{Diana, PacoTrex}.

Lastly, due to global-warming concerns new commercial gases are being introduced by industry to replace C$_2$H$_2$F$_4$, and it remains to be seen if they provide an advantage (generally, implying the drift velocity $v_d$ to be higher for the same working gain). In all, it seems very unlikely that the land-mark value of 10ps-$\sigma$ accuracy for mips will not be reached soon by a technology based on RPCs, even if the solution might be initially restricted to some specific environments. That's the scale at which delays on signal induction and particle transit due to the sheer size of the structure will start to matter. This represents an amazing situation by itself, and one that will require considerable ingenuity to be overcome.

\section*{Acknowledgements}

DGD wants to thank specially P. Fonte, N. Tomida and S. Duarte for discussions, and to the RD51 collaboration for support and encouragement.


\begin{thebibliography}{99}
\bibitem{Pestov}    V.V. Parkhomchuck, Yu. N. Pestov, N. V. Petrovykh, \textit{A spark counter with large area}, Nucl. Instr. Meth., 93(1971)269.
\bibitem{Santonico}	R. Santonico, R. Cardarelli, \textit{Development of Resistive Plate Counters}, Nucl. Instr. Meth. A, 187(1981)377.
\bibitem{Williams}  E. Cerron Zeballos et al., \textit{A new type of resistive plate chamber: The multigap RPC}, Nucl. Instr. Meth. A, 374(1996)132.
\bibitem{Fonte}     P. Fonte, A. Smirnitski, M.C.S. Williams, \textit{A new high-resolution TOF technology}, Nucl. Instr. Meth. A, 443(2000)201.
\bibitem{ALICE} A. Akindinov et al., \textit{Performance of the ALICE Time-Of-Flight detector at the LHC}, Eur. Phys. J. Plus 128(2013)44.
\bibitem{STAR}  W. J. Llope et al., \textit{Multigap RPCs in the STAR experiment at RHIC}, Nucl. Instr. Meth. A 661(2012)S110.
\bibitem{HADES} G. Kornakov et al., \textit{Time of flight measurement in heavy-ion collisions with the HADES RPC TOF wall}, JINST 9, 11(2014)C11015.
\bibitem{PET}   M. Couceiro et al., \textit{Spatial resolution of human RPC-PET system}, Nucl. Instr. Meth. A, 661, 1, (2012)156.
\bibitem{Neutrons} A. Blanco et al., \textit{Performance of timing resistive plate chambers with relativistic neutrons from 300 to 1500 MeV}, JINST 10, 2(2015)C02034.
\bibitem{Caba} P. Cabanelas, M. Morales et al., \textit{Performances of 4-gap timing RPCs for relativistic ions in the range $Z = 1$-$6$}, JINST 4(2009)P11007.
\bibitem{Para} C. Paradela et al., \textit{Performance of tRPC prototypes with relativistic heavy ions}, PoS(RPC2012)020.
\bibitem{DiegoIntegrity} D. Gonzalez-Diaz, Huangshan Chen, Yi Wang, \textit{Signal coupling and signal integrity in multi-strip resistive plate chambers used for timing applications}, Nucl. Instr. Meth. A, 648(2011)52.
\bibitem{FonteProp} P. Fonte, \textit{Frequency-domain formulation of signal propagation in multistrip Resistive Plate Chambers and its low-loss, weak-coupling analytical approximation}, JINST 8(2013)P08007.
\bibitem{R3B} M. Roder et al., \textit{Prototyping a 2-m x 0.5-m MRPC-based neutron TOF-wall with steel converter plates}, JINST 7(2012)P11030.
\bibitem{WangChinese} Jingbo Wang, Yi Wang et al., \textit{Development of multi-gap resistive plate chambers with low-resistive silicate glass electrodes for operation at high particle fluxes and large transported charges}, Nucl. Instr. Meth. A 621(2010)151.
\bibitem{Laso} A. Laso Garcia et al, \textit{High-rate timing resistive plate chambers with ceramic electrodes}, Nucl. Instr. Meth. A., 818(2016)45.
\bibitem{Fil}  S. Franchino at al.,  \textit{Effects of High Charge Densities in Multi-GEM Detectors} arXiv:1512.04968v1 [physics.ins-det].
\bibitem{CMS} J. Merlin, \textit{Study of long-term sustained operation of gaseous detectors for the high rate environment in CMS}, PhD. Thesis, Strasbourg 2016.
\bibitem{Sebastian} Sebastian White, \textit{Fast Timing R\&D for the HL-LHC Era}, CERN detector seminar, 25/10/2015. 
\bibitem{DiegoArchana} D. Gonzalez-Diaz, A. Sharma, \textit{Challenges for resistive gaseous detectors towards RPC2014}, JINST 8(2013)T02001.
\bibitem{PauloTime} A. Blanco, P. Fonte et al., \textit{Resistive plate chambers for time-of-flight measurements}, Nucl. Instr. Meth. A 513(2003)8.
\bibitem{DiegoPro} D. Gonzalez-Diaz, P. Fonte, J. A. Garzon, A. Mangiarotti, \textit{An analytical description of rate effects in timing RPCs},
Nucl. Phys. B (Proc. Suppl.) 158(2006)111.
\bibitem{JB} Jingbo Wang et al. \textit{Development of high-rate MRPCs for high resolution time-of-flight systems}, Nucl. Instr. Meth. A, 713(2013)40.
\bibitem{Repond} B. Bilki et al., \textit{Measurement of the rate capability of Resistive Plate Chambers}, JINST 4(2009)P06003.
\bibitem{Benciveni} G. Bencivenni et al., \textit{The micro-Resistive WELL detector: a compact spark-protected single amplification-stage MPGD}, JINST 1, 2(2015)P02008.
\bibitem{Urquijo} J. de Urquijo et al., \textit{Electron swarm coefficients in 1,1,1,2 tetrafluoroethane (R134a) and its mixtures with Ar}, Eur. Phys. J. D 51(2009)241.
\bibitem{CBM} I. Deppner et al., \textit{The CBM time-of-flight wall}, Nucl. Instr. Meth. A661(2012)S121.
\bibitem{Max} C. Adloff, M. Chefdeville et al., \textit{Resistive Micromegas for sampling calorimetry}, PoS(TIPP2014)054.
\bibitem{Charpak1} G. Charpak et al., \textit{Micromegas, a multipurpose gaseous detector} Nucl. Instr. Meth. A 478(2002)26.
\bibitem{ThomasLast} T. Papaevangelou et al., \textit{Fast Timing for High-Rate Environments with Micromegas}, arXiv:1601.00123v2 [physics.ins-det].
\bibitem{DiegoCurr}	D. Gonzalez-Diaz, \textit{RPC simulations from a current stand-point}, PoS(RPC2012)035.
\bibitem{FOPIandreas} A. Schuettauf, \textit{Timing RPCs in FOPI}, Nucl. Instr. Meth. A, 533(2004)65.
\bibitem{Clayton} C.R. Paul, \textit{Analysis of Multiconductor Transmission Lines}, John Wiley \& Sons, 2008.
\bibitem{Riegler} W. Riegler, D. Burgarth, \textit{Signal propagation, termination, crosstalk and losses in resistive plate chambers},
Nucl. Instr. Meth. A 481(2002)130.
\bibitem{DiegoFirst} D. Gonzalez-Diaz, \textit{Simulation of resistive plate chambers with multi-strip readout}, Nucl. Instr. Meth. A, 661(2012)172.
\bibitem{NatsukiProp} S. Tanaka et al., \textit{Development of a 2-m RPC for the LEPS2 Spectrometer}, at the XIII RPC Conference.
\bibitem{NatsukiProp2} N. Tomida et al., \textit{The basic performance of the TOF-RPC for the BGOegg experiment}, at the XIII RPC Conference.
\bibitem{Natsuki} Courtesy of Natsuki Tomida.
\bibitem{STARHS}  Huangshan Chen et al., \textit{R\&D and mass production of LMRPC modules for the STAR-MTD system}, JINST 7(2012)P10006.
\bibitem{ItalianLady} B. Liberti, \textit{Further gas mixtures with low enviroment impact}, at the XIII RPC Conference.
\bibitem{NINO} F. Anghinolfi et al., \textit{NINO: an ultra-fast and low-power front-end amplifier/discriminator ASIC designed for the multigap resistive plate chamber}, Nucl. Instr. Meth. 533(2004)183.
\bibitem{PADI} M. Ciobanu et al., \textit{PADI, an Ultrafast Preamplifier - Discriminator ASIC for Time-of-Flight Measurements}. IEEE Trans. Nucl. Sci. 61 (2014) no.2, 1015.
\bibitem{BendModel} R. Garg, I. J. Bahl, \textit{Microstrip discontinuities.} Int. J. Electronics, 45, 1(1978)81.
\bibitem{Allegro} http://www.cadence.com/products/pcb/ams\_simulator/pages/default.aspx, last accessed on 05/06/2016.
\bibitem{Williams20} S. An et al., \textit{A 20ps timing device -A Multigap Resistive Plate Chamber with 24 gas gaps}, Nucl. Instr. Meth. A 594(2008)39.
\bibitem{Babkin} V. Babkin, \textit{Triple-stack multigap resistive plate chamber with strip readout}, Nucl. Instr. Meth. A 824(2016)490.
\bibitem{PacoMicroBulk} F. J. Iguaz et al., \textit{Characterization of microbulk detectors in argon- and neon-based mixtures}, JINST 7(2012)P04007.
\bibitem{Elisa} E. Ruiz-Choliz et al., \textit{Modelling the behaviour of microbulk Micromegas in Xenon/trimethylamine gas}, Nucl. Instr. Meth. A 799(2015)137.
\bibitem{WernerChristian} C. Lippmann, W. Riegler, \textit{Space charge effects in Resistive Plate Chambers}, Nucl. Instr. Meth. A 517(2004)54.
\bibitem{Diana} S. Cebrian et al., \textit{Micromegas-TPC operation at high pressure in xenon-trimethylamine mixtures}, JINST 8(2013)P01012.
\bibitem{PacoTrex} F.J. Iguaz et al., \textit{TREX-DM: a low-background Micromegas-based TPC for low-mass WIMP detection} https://arxiv.org/pdf/1512.01455v1.pdf.
\end{thebibliography}
\end{document}